\documentclass[11pt]{article}
\usepackage{amsmath,amssymb,color,graphics,epsfig,cite}

\usepackage{amsfonts}

\newcommand{\be}{\begin{equation}}
	\newcommand{\ee}{\end{equation}}
\newcommand{\bea}{\setlength\arraycolsep{2pt} \begin{eqnarray}}
	\newcommand{\eea}{\end{eqnarray}}
\newcommand{\nn}{\nonumber}

\def\ft#1#2{{\textstyle{\frac{\scriptstyle #1}{\scriptstyle #2} } }}
\def\fft#1#2{{\frac{#1}{#2}}}

\def\0{{\sst{(0)}}}
\def\1{{\sst{(1)}}}
\def\2{{\sst{(2)}}}
\def\3{{\sst{(3)}}}
\def\4{{\sst{(4)}}}
\def\5{{\sst{(5)}}}
\def\6{{\sst{(6)}}}
\def\7{{\sst{(7)}}}
\def\8{{\sst{(8)}}}
\def\sst#1{{\scriptscriptstyle #1}}

\begin{document}
	
\begin{center}

{\Large {\bf Odd-dimensional Extremal Rotating Black Holes with All Equal Angular Momenta and Small Electric Charges}}

\vspace{20pt}

Qi-Yuan Mao and H. L\"u
		
\vspace{10pt}
		
{\it Center for Joint Quantum Studies, Department of Physics,\\
			School of Science, Tianjin University, Tianjin 300350, China }

\vspace{40pt}

\underline{ABSTRACT}

\end{center}
	
We consider Einstein-Maxwell gravity in diverse dimensions and construct the small charge perturbation to the extremal rotating black holes with all equal angular momenta in odd $D=2n+1$ dimensions. Exact solutions exist at the next-to-leading order (NLO), and they are analytic, allowing us to obtain the charge corrections to thermodynamic quantities at this order. Irrational exponents in the near-horizon power-series expansion emerge at the next-to-next-to-leading order (NNLO). We show, by numerical computation, that these horizon geometries can indeed be integrated out to asymptotic Minkowski spacetime, thereby proving the existence of the unusual singular horizon behavior of the extremal charged rotating black holes.

\vfill{qiyuan\_mao@tju.edu.cn \ \ \  mrhonglu@gmail.com}

	
\thispagestyle{empty}
\pagebreak

\newpage

\section{Introduction}

Electromagnetism and gravity are two known long-range forces in our universe, and they can be described by Einstein-Maxwell (EM) gravity. Despite its high nonlinearity, EM theory admits a variety of important exact solutions, including the Reissner-Nordstr\"om (RN) black hole, Kerr-Newman (KN) black hole \cite{Newman:1965my}, and their asymptotic AdS counterparts \cite{Carter:1968ks,Carter:1973rla}, all culminated into the Demianski-Plebanski solution \cite{Plebanski:1976gy} of the type D in Petrov classification. Both the generalizations of RN black hole or the neutral rotating black holes to higher dimensions are known, without \cite{Myers:1986un} or with a cosmological constant \cite{Hawking:1998kw,Gibbons:2004uw, Gibbons:2004js}. However, there are no known exact solutions of charged rotating black holes in higher dimensions, except those in supergravities, e.g.~\cite{Cvetic:1996xz,Chong:2005hr,Wu:2011gq}.

There are two corners of parameter space where one may explore the construction of perturbative
solutions of charged rotating black holes in higher dimensions. One is to consider slowly rotating solution \cite{Aliev:2005npa} and the other is for small charges.  In the former case, one starts with an RN black hole, and owing to the no-force condition in the extremal limit, the solutions are analytic and a small perturbation by rotation will not alter this fact. Furthermore, RN black hole with large amount charges tend to be more theoretical since they are unlikely to exist in our Universe. The latter situation is more realistic, since celestial bodies tend to have large spin with small charges, if there are any at all.

Of course, searching for a realistic black hole in higher dimensions is not a well-informed motivation. The latter case is worth studying for its own sake, since it is much subtler in the extremal limit, which balances attractive gravity and the repulsive centrifugal force. The balance is unstable, and a perturbation may create unusual structure of the near-horizon geometry. In \cite{Mao:2023qxq}, quadratic curvature perturbation of the extremal rotating black holes with all equal angular momenta in odd dimensions was studied, and it was observed that the near horizon geometry involves irrational exponents in the power series expansion, i.e. $(r-r_0)^{\Delta_+}$ where $\Delta_+$ is an irrational number, and $r_0$ is the radial location of the horizon. Despite the fact that all invariant polynomials of the curvature tensor and their covariant derivatives are regular on the horizon, the horizon is singular, giving rise to an impenetrable natural boundary of the spacetime, owing to the irrationality \cite{Mao:2023qxq}. At the first sight, one might attribute such unusual structure to higher-derivative theories, but it was argued in \cite{Mao:2023qxq} that even in (two-derivative) Einstein gravity, a perturbation from the usual matter energy-momentum tensor may also lead to such irrational exponents. Thus, extremal rotating black holes in Einstein-Maxwell theory carrying small charges become a simple but nontrivial testing ground of this idea.

Irrational exponents in the near-horizon geometries of extremal black holes arising from
perturbation are widespread phenomena \cite{Horowitz:2022mly,Horowitz:2023xyl,Horowitz:2024dch}.  However, we should be cautious that not all horizon geometries necessarily lead to black holes, especially in asymptotically-flat spacetimes where the no-hair theorem is much more stringent than the asymptotically-locally AdS spacetimes. For example, horizon geometries carrying independent local scalar hair exist in Einstein theory minimally coupled to a free scalar, but the no-hair theorem excludes the possibility that such a near-horizon geometry can be smoothly connected to asymptotic Minkowski spacetime. We therefore believe that it is insufficient to study black hole property by analysing the near-horizon geometry alone. In this paper, we shall not only examine whether the near-horizon geometries with irrational exponents can arise, but also verify whether they can be smoothly connected to asymptotic infinity.

The construction of charged rotating black holes in general dimensions are complicated, even at the perturbative level, but this is a must procedure in order to test whether a solution with irrational exponents truly describes a black hole. We circumvent the difficulty of the construction by considering odd dimensional black holes with all equal angular momenta. In $D=2n+1$ dimensions, there can be $n$ number of orthogonal rotations, giving rise to $n$ independent angular momenta. The metric is cohomogeneity-$n$ with only $(n+1)$ Killing vectors. When the angular momenta are all equal, the metric reduces to cohomogeneity-one with one time-like Killing vector and level surfaces of squashed $S^{2n-1}$ described as a $U(1)$ bundle over $\mathbb{CP}^{n-1}$. In this paper, we use the Ricci-flat Myers-Perry (MP) black hole in $D=2n+1$ dimensions with all equal angular momenta as the leading-order solution and obtain the general NLO charge perturbation. We then use $D=7$ as an example to construct the NNLO solution in the charge perturbation and show that irrational exponents indeed emerge in these extremal black holes.

The paper is organized as follows. In Section 2, we consider EM gravity and present the ansatz for charged rotating black holes with all equal angular momenta in odd dimensions. The ansatz involving six functions and we derive their nonlinear ordinary differential equations of motion and present them in Appendix A. In Section 3, we consider the perturbative approach and obtain the exact NLO solution, and study the small charge corrections to the thermodynamic quantities. In Section 4, we continue to the NNLO in $D=7$, and obtain the numerical solution, from which we obtain the NNLO corrections to the black hole thermodynamic quantities. We analyse the horizon geometry with irrational exponents. We conclude the paper in Section 5.

\section{Ansatz}
	
In this paper, we consider EM gravity in diverse dimensions. The Lagrangian is given by
\begin{equation}
{\cal L}=\sqrt{-g} \left(R - \ft{1}{4} F_{\mu\nu}F^{\mu\nu}\right)\,,\qquad
F_{\mu\nu}=\partial_\mu A_\nu - \partial_\nu A_\mu\,.
\end{equation}
The Einstein and Maxwell field equations are
\begin{equation}
R_{\mu\nu} - \ft12 R g_{\mu\nu} = \ft12 (F_{\mu\rho}F^\rho{}_{\nu} - \ft{1}{4} g_{\mu\nu}F_{\rho\sigma}F^{\rho\sigma})\,, \qquad \nabla_{\mu}F^{\mu\nu} = 0 \,.
\label{covariant tensor eom}
\end{equation}
As was mentioned in Introduction, the theory admits a variety of charged solution in four dimensions, but the generalization to higher dimensions meets limited success. In this paper, we consider charged extremal rotating black holes. For simplicity, we assume that all angular momenta are equal and work in odd $D=2n+1$ dimensions. The resulting geometry is cohomogeneity-one, with the ansatz
\begin{eqnarray}
ds_{2n+1}^2 &=&-\frac{h(r)}{W(r)}dt^2+\frac{dr^2}{f(r)}+r^2W(r)(\sigma+\omega(r)dt)^2
+r^2ds_{\mathbb{CP}^{n-1}}^2 \,,\nn\\
A &=& \Psi(r)\, dt + \Phi(r)\, \sigma \,.\label{metric-unpertur-D}
\end{eqnarray}
Here, $ds^2_{\mathbb{CP}^{n-1}}$ is the metric of a $2(n-1)$-dimensional complex projective space, and $\sigma=d\psi + {\cal A}$ is the 1-form connection along the Hopf fiber direction. The coordinate $\psi$ has period $2\pi$ and $d{\cal A}=2J$, where $J$ is the K\"ahler 2-form on $\mathbb{CP}^{n-1}$. The combination $(\sigma^2 +ds^2_{\mathbb{CP}^{n-1}})$ describes the metric of unit round $S^{2n-1}$, with $R^{i}{}_j=2(n-1) \delta^i_j$ \cite{Gibbons:2004ai}.

The ansatz involves six functions, $(h,f,W,\omega, \Psi,\Phi)$. Substituting the ansatz into the covariant equations of motion in \eqref{covariant tensor eom} yields a set of second-order nonlinear ordinary differential equations, which we present in Appendix \ref{app:gen-eom}. We do not expect that exact solutions exist in general $D$ dimensions, and we shall consider a perturbative approach with the electric charge as the perturbation parameter.

\section{Perturbative approach and general NLO solution}

The general analytic solutions to equations in \eqref{gen-eom} are unlikely to exist. We consider small charge perturbation to the neutral Ricci-flat MP solution. In other words, we consider the MP solution with all equal angular momenta as the leading-order solution and construct the perturbative solutions order by order, with the electric charge as the order parameter.

We adopt the notation of \cite{Feng:2016dbw}, and the leading-order Ricci-flat solution is given by
\begin{equation}
\bar{W}=1+\frac{\nu^2}{r^{D-1}}\,,\qquad\bar{\omega}(r)=\frac{\sqrt\mu\nu}{r^{D-1}\bar{W}}\,,
\qquad\bar{f}=\bar{h}=1-\frac{\mu}{r^{D-3}}+\frac{\nu^2}{r^{D-1}} \,. \label{ansatz-unpertur-D}
\end{equation}
The horizon of the black hole is located at $\bar f(r_0)=0$, and the complete set of black hole thermodynamic quantities are
\begin{eqnarray}
M_0&=&\frac{(D-2)\Omega_{D-2}}{16\pi}\mu\,,\qquad J_0=\frac{(D-1)\Omega_{D-2}}{16\pi}\sqrt{\mu}\nu\,,\qquad\Omega_0=
\frac{\nu}{r_0\sqrt{r_0^{D-1}+\nu^2}}\,,\cr
T_{0}&=&\frac{(D-3)r_{0}^{D-1}-2\nu^{2}}{4\pi r_{0}^{\frac{1}{2}(D+1)}\sqrt{r_{0}^{D-1}+\nu^{2}}}\,,\qquad S_{0}=\frac{\Omega_{D-2}}{4}r_{0}^{\frac{1}{2}(D-3)}\sqrt{r_{0}^{D-1}+\nu^{2}} \,. \label{lv0-thermo}
\end{eqnarray}
It can be easily verified that they satisfy the first law of black hole thermodynamics. In the extremal limit, where $T_0=0$, we have
\begin{equation}
\mu = \frac{D-1}{2 r_0^{D-3}}\,, \qquad \nu = \sqrt{\frac{D-3}{2}} r_0^{\frac{D-1}{2}}\,.\label{munur0}
\end{equation}
We consider perturbations of the extremal black holes, up to and including the NNLO. Considering the fact that Maxwell equation is linear, but the energy-momentum tensor is quadratic in Maxwell field, the perturbative Maxwell field is given by
\begin{eqnarray}
\Psi = q \delta_{q^1}\psi +q^3 \delta_{q^3}\psi \,,\qquad
\Phi = q \delta_{q^1}\phi + q^3\delta_{q^3}\phi \,, \label{lv2-A-ansatz}
\end{eqnarray}
where we set the charge parameter $q$ as the perturbative parameter and omitted the ${\cal O}(q^5)$ symbol. The feedback of this perturbation to the metric creates the $q^2$ and $q^4$ orders in perturbation:
\begin{eqnarray}
W &=&\bar W + q^2 \delta_{q^2}W + q^4 \delta_{q^4}W \,,\qquad
f = \bar f\left(1 + q^2 \delta_{q^2}f + q^4\delta_{q^4}f\right) \,,\cr
h &=& \bar h\left(1 + q^2 \delta_{q^2}h + q^4\delta_{q^4}h\right) \,, \qquad
\omega = \frac{\sqrt[]{\mu} \nu}{r^{D-1} W}+q^2 \delta_{q^2}\omega(r) + q^4\delta_{q^4}\omega \,. \label{lv2-g-ansatz}
\end{eqnarray}
In other words, the metric functions are in even powers of $q$ in the perturbative expansions whilst the Maxwell fields are in odd powers. Note that in this perturbative approach, the horizon position $r_+=r_0$ remains unchanged, given by \eqref{munur0}. We adopt this horizon-fixed perturbation scheme for the later convenience of numerical analysis, where we need to solve the differential equations from the horizon $r_0$ to asymptotic infinity.

It turns out that the NLO solution can be solved analytically. We find the NLO solution to the Maxwell equation is
\begin{equation}
\delta_{q^1} \psi(r) = \frac{c}{r^{D-3}} \,,\qquad
\delta_{q^1} \phi(r) = \frac{c \sqrt{D-3}}{\sqrt{D-1}\, r^{D-3}} \,,
\end{equation}
where $c$ is an integration constant that can be absorbed into the perturbation parameter $q$. We therefore set $c=1$ without loss of generality. Consequently, the $q^2$-order back reaction to the metric functions are given by
\begin{eqnarray}
\delta_{q^2} f &=& \frac{(D-3) \left(\left((D-5) \rho ^2-D+1\right) \rho ^D+\rho ^3 \left((D-1) \rho ^2-D+5\right)\right)}{(D-2) (D-1) r_0^{2(D-3)}\rho^{D-1} \left(2 \rho ^D-(D-1) \rho ^3+(D-3) \rho \right)}=\delta_{q^2} h\,,\nn\\
\delta_{q^2} W &=& \frac{(D-3) \left((D-1) \rho ^D+(D-3) \rho ^3\right)}{2 (D-2) (D-1) \rho ^{2 D-1} r_0^{2(D-3)}}\,,\nn\\
\delta_{q^2} \omega &=& -\frac{\left(\frac{D-3}{D-1}\right)^{3/2} \left((D-2) \rho ^D+(D-1) \rho ^3\right)}{(D-2) \left(2 \rho ^D+(D-3) \rho \right)r_0^{2 D-5} \rho ^{D-1}}\,,
\end{eqnarray}
where $\rho=r/r_0$ is the dimensionless radius. In the above, we have chosen all the integration constants appropriately so that the metric is asymptotically-flat with no rotations.

Having obtained the NLO perturbative solution, we can use the standard technique to derive all the perturbed thermodynamic quantities to this order. We find
\begin{eqnarray}
M&=& \frac{(D-2) (D-1)r_0^{D-3}}{32 \pi } \Big(1-\frac{(D-5) (D-3)q^2}{(D-2) (D-1)^2 r_0^{2 (D-3)}}\Big)\Omega_{D-2}\,,\nn\\
J&=& \frac{(D-1) \sqrt{(D-3) (D-1)} r_0^{D-2}}{32 \pi }\Big(1-\frac{(D-3)q^2 }{(D-1)^2 r_0^{2 (D-3)}}\Big)\Omega_{D-2}\,,\nn\\
Q&=& \frac{(D-3)q}{16 \pi }  \Omega_{D-2}\,,\qquad
\Omega = \frac{\sqrt{\frac{D-3}{D-1}}}{r_0}\Big(1+\frac{(D-3)q^2 }{(D-2) (D-1)^2 r_0^{2(D -3)}}\Big)\,,\nn\\
\Phi &=& \frac{2q}{(D-1) r_0^{D-3}}\,,\qquad
S = \frac{\sqrt{D-1}}{4 \sqrt{2}} r_0^{D-2} \left( 1-\frac{(D-3) q^2}{ (D-1)^{2} r_0^{2(D-3)}}\right)\Omega _{D-2}\,.
\end{eqnarray}
Note that we have omitted the ${\cal O}(q^3)$ and ${\cal O}(q^4)$ notations in the above expressions. It is easy to verify that the first law of the extremal black hole
\be
dM=\Omega dJ + \Phi dQ\,,\label{firstlaw}
\ee
is satisfied up to and including the NLO. The mass and entropy now are functions of both $J$ and $Q$, given by
\bea
M &=& \eta_0\Big(J^{\fft{D-3}{D-2}} + \eta_2 Q^2 J^{-\fft{D-3}{D-2}}\Big) + {\cal O}(Q^4)\,,\qquad
S = \frac{4 \sqrt{2} \pi }{\sqrt{D-3} (D-1)} J + {\cal O}(Q^4)\,,\nn\\
\eta_0 &=& \frac{(D-2)\left(\frac{\Omega_{D-2} }{32 \pi }\right)^{\frac{1}{D-2}}}{(D-3)^{\frac{D-3}{2 (D-2)}} (D-1)^{\frac{D-5}{2 (D-2)}}}\,,\qquad
\eta_2 = \frac{(D-3)^{\frac{1}{2-D}} (D-1)^{\frac{D-5}{D-2}}}{2 (D-2)  \left(\frac{\Omega }{32 \pi }\right)^{\frac{2}{D-2}}}\,.\label{NLOmasscharge}
\eea
Note that the small charge contributes positively to the mass, while the entropy/charge relation receives no correction at the $Q^2$ order.

\section{The NNLO solution in $D=7$}

In the previous section, we obtained the NLO solution in general odd dimensions. We find that at this order, perturbations can be solved exactly and the solutions are analytic functions, instead of having irrational exponents. This appears to be in contradiction to our expectation. We proceed to compute the NNLO solution and find that the equations cannot be all solved analytically. We shall consider $D=7$ as an illustrative example, as in the case of the higher-derivative correction in \cite{Mao:2023qxq}.

\subsection{Maxwell equation and the solution}

At the NNLO, the equation of motion for the Maxwell field $A$ becomes
\begin{eqnarray}
\frac{64}{3} r_0^3&=& 5 r^5 r_0^5 \big(5 r^6-2 r_0^6\big) \delta_{q^3}\psi'(r) + 5 \sqrt{6} r^5 r_0^{10} \delta_{q^3}\phi'(r) +5 r^{12} r_0^5 \delta_{q^3}\psi''(r)\nn\\
&&+  10 r^6 r_0^{11} \delta_{q^3}\psi''(r) - 5 \sqrt{6} r^6 r_0^{10} \delta_{q^3}\phi''(r) \,,\nn \\
-8 \sqrt{6} &=& 40 r^8 r_0^3 \delta_{q^3}\phi(r) + 5 \sqrt{6} r^5 r_0^8 \delta_{q^3}\psi'(r)-15 r^9 r_0^3 \delta_{q^3}\phi'(r)-15 r^5 r_0^7 \delta_{q^3}\phi'(r)\nn\\
&&-5 \sqrt{6} r^6 r_0^8 \delta_{q^3}\psi''(r)-5 r^{10} r_0^3 \delta_{q^3}\phi''(r)+15 r^6 r_0^7 \delta_{q^3}\phi''(r) \,.
	\end{eqnarray}
These are two second-order linear differential equations with sources (from the NLO) appearing in the left-hand side of the equations. It turns out that they can be both analytically solved, yielding
\begin{eqnarray}
\delta_{q^3}\psi &=&\frac{1}{450r_0^{12} r^8}\Big(-6 r_0^2 \big(r^2+r_0^2\big) \big(-13 r^4+6 r^2 r_0^2+3 r_0^4\big)\nn\\
&&+r^4 \big(r^4-r_0^4\big) \Big(-162 \log (r)+80 \log \big(r^2-r_0^2\big)+\log \big(r^2+2 r_0^2\big)\Big)\Big),\nn \\
\delta_{q^3}\phi &=&\frac{1}{225\sqrt{6}r_0^{13}r^8}\Big(2 r_0^2 \big(39 r^8+21 r^6 r_0^2+r^4 r_0^4-27 r^2 r_0^6-9 r_0^8\big)\nn\\
&&+r^4 \big(r^6-r_0^6\big) \Big(-162 \log (r)+80 \log \big(r^2-r_0^2\big)+\log \big(r^2+2 r_0^2\big)\Big)\Big).\label{q4maxsource}
\end{eqnarray}
Although the solutions involve logarithmic terms, the solutions are finite in the neighborhood of $(r-r_0)$ and $r\rightarrow\infty$. We have chosen the integration constants  so that the solutions can be viewed as induced by the sources only, with the sourceless components removed. It reflects the fact that $\delta_{q^3}\psi$ has the ${1}/{r^{10}}$ leading-order falloff at asymptotic infinity instead of the ${1}/{r^4}$ behavior.

\subsection{Einstein equations}

The Einstein equations of motion  at the NNLO $\mathcal{O}(q^4)$ order are much more complicated.	Through the successive elimination of variables in the differential equations, the system of the coupled differential equations of the four metric functions can be transformed into one decoupled fourth-order differential equation of the metric function $f$, given by
\begin{eqnarray}
P_4 \delta_{q^4}f''''+ P_3 \delta_{q^4}f''' + P_2 \delta_{q^4}f'' + P_1 \delta_{q^4}f' + P_0 \delta_{q^4}f = Q \,, \label{f-eom-q4}
\end{eqnarray}
where
\begin{eqnarray}
P_4 &=&125 r^{12} r_0^{12} \left(r^4+r^2 r_0^2-2 r_0^4\right){}^4 \left(9 r^6+18 r^4 r_0^2+17 r^2 r_0^4+16 r_0^6\right)\,,\cr
P_3 &=& 500 r^{11} r_0^{12} \left(r^4+r^2 r_0^2-2 r_0^4\right){}^3 \big(54 r^{10}+162 r^8 r_0^2+281 r^6 r_0^4+330 r^4 r_0^6\cr
&&+189 r^2 r_0^8+64 r_0^{10}\big)\,,\nn \\
P_2 &=& 375 r^{10} r_0^{12} \left(r^4+r^2 r_0^2-2 r_0^4\right){}^2 \big(435 r^{14}+1740 r^{12} r_0^2+3734 r^{10} r_0^4\nn\\
&&+5708 r^8 r_0^6+5979 r^6 r_0^8+3920 r^4 r_0^{10}+1796 r^2 r_0^{12}+448 r_0^{14}\big)\,,\nn \\
P_1 &=& 375 r^9 \left(r-r_0\right) r_0^{12} \left(r+r_0\right) 	\left(r^2+r_0^2\right) \left(r^2+2 r_0^2\right){}^2 \big(285 r^{14}+570 r^{12} r_0^2\nn\\
&&+1308 r^{10} r_0^4+1422 r^8 r_0^6+1523 r^6 r_0^8+2184 r^4 r_0^{10}+900 r^2 r_0^{12}+448 r_0^{14}\big)\,,\nn \\
P_0 &=& -12000 r^{14} r_0^{12} \left(r^2+2 r_0^2\right){}^2 \big(45 r^{12}+90 r^{10} r_0^2+149 r^8 r_0^4+136 r^6 r_0^6\nn\\
&&+48 r^4 r_0^8+40 r^2 r_0^{10}+32 r_0^{12}\big)\,,\label{Pi}
\eea
together with a source term
\bea
Q &=& \frac{64}{3} r_0^4 \Big(22 r^4 \big(r-r_0\big) \big(r+r_0\big) \big(r^2+2 r_0^2\big){}^2 \big(15 r^{12}+45 r^{10} r_0^2\nn\\
&&+284 r^8 r_0^4+189 r^6 r_0^6-181 r^4 r_0^8-156 r^2 r_0^{10}-256 r_0^{12}\big)\nn\\
&&-2 \big(165 r^{22}+405 r^{20} r_0^2+784 r^{18} r_0^4-7296 r^{16} r_0^6-47672 r^{14} r_0^8\nn\\
&&-58060 r^{12} r_0^{10}+65355 r^{10} r_0^{12}+162151 r^8 r_0^{14}+103616 r^6 r_0^{16}\nn\\
&&+18312 r^4 r_0^{18}-56960 r^2 r_0^{20}-51200 r_0^{22}\big)-r^4 \big(r-r_0\big) \nn\\
&&\big(r+r_0\big) \big(r^2+2 r_0^2\big){}^2 \big(15 r^{12}+45 r^{10} r_0^2+284 r^8 r_0^4+189 r^6 r_0^6-181 r^4 r_0^8\nn\\
&&-156 r^2 r_0^{10}-256 r_0^{12}\big) \Big(162 \log (r)-80 \log \big(r^2-r_0^2\big)-\log \big(r^2+2 r_0^2\big)\Big)\Big).
\end{eqnarray}
It is worth pointing out that the polynomials $P_i's$ are determined from the Einstein tensor, while the source comes from the contribution of the Maxwell energy-momentum tensor. Analogous equation was obtained in \cite{Mao:2023qxq} in the study of higher-order curvature perturbation to the Kerr black hole. In particular, the $P_i's$ are identical, while the source came from the higher-derivative contribution in \cite{Mao:2023qxq}.

If we can solve for $\delta_{q^4} f$ in \eqref{f-eom-q4}, the remaining perturbed metric functions can be successively obtained, given by
\begin{eqnarray}
		\delta_{q^4}W &=& \frac{1}{13500 r^{16} r_0^{12} \big(9 r^6+18 r^4 r_0^2+17 r^2 r_0^4+16 r_0^6\big)}\Big(-16 r_0^2 \big(1755 r^{16}\cr
		&&+4455 r^{14} r_0^2+4086 r^{12} r_0^4+2016 r^{10} r_0^6-5002 r^8 r_0^8-5861 r^6 r_0^{10}\cr
		&&+6219 r^4 r_0^{12}+3032 r^2 r_0^{14}-3200 r_0^{16}\big)+3 r^4 \big(r-r_0\big) \big(r+r_0\big) \Big(-9000 r^6 r_0^{12} \cr
		&&\big(8 r^{10}+24 r^8 r_0^2+38 r^6 r_0^4+43 r^4 r_0^6+33 r^2 r_0^8+16 r_0^{10}\big) \delta_{q^4}f(r)\cr
		&&+8 \big(15 r^{12}+45 r^{10} r_0^2+60 r^8 r_0^4+57 r^6 r_0^6-r^4 r_0^8-28 r^2 r_0^{10}+32 r_0^{12}\big)\cr
		&& \log \Big(\frac{\big(r^2-r_0^2\big){}^{80} \big(r^2+2 r_0^2\big)}{r^{162}}\Big)+375 r^5 r_0^{12} \Big(-2 \big(r^2-r_0^2\big){}^2 \big(4 r^8+20 r^6 r_0^2\cr
		&&+45 r^4 r_0^4+56 r^2 r_0^6+28 r_0^8\big) \delta_{q^4}f'(r)+r \big(r^4+r^2 r_0^2-2 r_0^4\big){}^2 \cr
		&&\Big(14 \big(r^4+r^2 r_0^2+r_0^4\big) \delta_{q^4}f''(r)+r \big(r^4+r^2 r_0^2-2 r_0^4\big) \delta_{q^4}f^{(3)}(r)\Big)\Big)\Big)\Big)\,,\cr
		\delta_{q^4}h' &=& \frac{1}{675 \big(5 r^6+4 r_0^6\big)}\Big(\frac{16}{r^9 r_0^{12}}\Big(-\frac{2 r_0^2}{r^2+2 r_0^2}\big(39 r^{10}+99 r^8 r_0^2+76 r^6 r_0^4\nn\\
		&&+158 r^4 r_0^6+198 r^2 r_0^8+72 r_0^{10}\big)+r^4 \big(r^6+2 r_0^6\big) \log \Big(\frac{\big(r^2-r_0^2\big){}^{80} \big(r^2+2 r_0^2\big)}{r^{162}}\Big)\Big)\nn\\
		&&+675 \Big(\big(5 r^6+4 r_0^6\big) \delta_{q^4}f'(r)+r^6 \Big(7 \delta_{q^4}W'(r)+r \delta_{q^4}W''(r)\Big)\Big)\Big)\,,\nn\\
		\delta_{q^4}\omega' &=& \frac{1}{4050 \sqrt{6} r^{11} r_0^{17} \big(r^6+2 r_0^6\big){}^3} 4 r_0^2 \big(312 r^{26}+168 r^{24} r_0^2-265 r^{22} r_0^4\nn\\
		&& +1872 r^{20} r_0^6+966 r^{18} r_0^8+1398 r^{16} r_0^{10}+3744 r^{14} r_0^{12}+10116 r^{12} r_0^{14}\nn\\
		&&-3936 r^{10} r_0^{16}+6816 r^8 r_0^{18}+5808 r^6 r_0^{20}-2624 r^4 r_0^{22}+3456 r^2 r_0^{24}+384 r_0^{26}\big)\nn\\
		&&+2700 r^8 r_0^{12} \big(5 r^6-2 r_0^6\big) \big(r^6+2 r_0^6\big){}^3 \delta_{q^4}f(r)-72900 r^{16} r_0^{22} \big(r^6+2 r_0^6\big) \delta_{q^4}h(r)\nn\\
		&&+r^4 \big(r^6+2 r_0^6\big) \Big(2700 r^{10} r_0^{12} \big(r^{12}+r^6 r_0^6+27 r^2 r_0^{10}+16 r_0^{12}\big) \delta_{q^4}W(r)\nn\\
		&&+\big(r^6+2 r_0^6\big) \Big(8 \big(r-r_0\big) \big(r+r_0\big) \big(2 r^4+2 r^2 r_0^2-r_0^4\big) \big(r^6+2 r_0^6\big)\nn\\
		&& \log \Big(\frac{\big(r^2-r_0^2\big){}^{80} \big(r^2+2 r_0^2\big)}{r^{162}}\Big)+675 r^5 r_0^{12} \Big(\big(r-r_0\big){}^2 \big(r+r_0\big){}^2 \big(r^2+2 r_0^2\big)\nn\\
		&& \big(5 r^6+4 r_0^6\big) \delta_{q^4}h'(r)+r^6 \big(-r^6+9 r^2 r_0^4+4 r_0^6\big) \delta_{q^4}W'(r)\Big)\Big)\Big)\,.\label{eom on omega}
\end{eqnarray}

\subsection{Asymptotic Behavior}

In order to study the properties of black hole at the NNLO, it is necessary to solve the equations in \eqref{f-eom-q4} and \eqref{eom on omega}. We first determine the behavior of the function $\delta_{q^4}f$ at the asymptotic infinity.  We define
\begin{equation}
\delta_{q^4}f = r^ \lambda \delta_{q^4}f_{\text{inf}}\,. \label{f_inf-def}
\end{equation}
Substituting this into \eqref{f-eom-q4} and taking the $r\rightarrow \infty$ limit, we find
\begin{equation}
1125r_0^{2}r^{\lambda+14}(-2+\lambda)(4+\lambda)(6+\lambda)(10+\lambda) \delta_{q^4}f_{\text{inf}} = -13824 \,. \label{f_inf_eq}
\end{equation}
Therefore, we have $\lambda = 2,-4,-6,-10$ for the source-free contributions and $\lambda = -14$ for the source contribution. The general asymptotic solution can be written as
	\begin{equation}
		\delta_{q^4}f = -\frac{3}{1250r_0^{2}r^{14}} \tilde{f}_{0}(r) + \frac{C_{10}}{r_0^6 r^{10}}\tilde{f}_{10}(r) + \frac{C_{6}}{r_0^{10}r^6}\tilde{f}_{6}(r) + \frac{C_{4}}{r_0^{12} r^4}\tilde{f}_{4}(r) + \frac{C_{-2} r^2}{r_0^{18}}\tilde{f}_{-2}(r) \,. \label{f-inf-ansatz}
	\end{equation}
	Here $\tilde{f}_{i}$'s all take the form $\tilde{f}_{i} = 1+ \frac{A_1}{r} + \frac{A_2}{r^2} + \cdots$, where $A_i$'s denote constants. The integration constants $C_i$'s are scaled such that they are dimensionless numbers. Having four independent integration constants indicates that \eqref{f-inf-ansatz} accounts for the most general solution for the fourth-order differential equation \eqref{f-eom-q4}.

The requirement that the asymptotic infinity should be Minkowskian implies that we must set $C_{-2}=0$. Therefore, the general asymptotically-flat perturbations involve three independent free integration constants $(C_4,C_6,C_{10})$. We can also determine the $\tilde f_i$ order by order in the large-$r$ expansion. We give the results for the low-lying falloff orders:
\begin{eqnarray}
\tilde{f}_{0}(r) &=&  1-\frac{352 r_0^2}{81 r^2}+\frac{10037 r_0^4}{756 r^4}-\frac{58654 r_0^6}{2079 r^6}+\frac{5521955 r_0^8}{72576 r^8} + \cdots\,,\nn \\
\tilde{f}_{4}(r) &=& 1+\frac{3 r_0^4}{r^4}+\frac{9 r_0^8}{r^8}-\frac{177 r_0^{10}}{80 r^{10}}+\frac{127 r_0^{12}}{5 r^{12}}-\frac{5289 r_0^{14}}{320 r^{14}}+\frac{44829 r_0^{16}}{616 r^{16}} +\cdots\,,\nn\\
\tilde{f}_{6}(r) &=& 1-\frac{2 r_0^6}{r^6}-\frac{909 r_0^8}{160 r^8}-\frac{18 r_0^{10}}{5 r^{10}}-\frac{16133 r_0^{12}}{640 r^{12}}+\frac{15201 r_0^{14}}{1232 r^{14}}-\frac{2038257 r_0^{16}}{20480 r^{16}}+\cdots\,,\nn \\
\tilde{f}_{10}(r) &=& 1+\frac{783 r_0^4}{160 r^4}-\frac{14 r_0^6}{5 r^6}+\frac{11991 r_0^8}{640 r^8}-\frac{27243 r_0^{10}}{1232 r^{10}}+\frac{1478139 r_0^{12}}{20480 r^{12}}+\cdots\,.
\end{eqnarray}
The large-$r$ expansion of the other metric functions can also be determined, given by
\begin{eqnarray}
\delta_{q^4}\tilde{W}  &=& \big(\frac{44}{3375 r^{10} r_0^6}+\frac{176}{16875 r^{16}}-\frac{71}{2250 r^{14} r_0^2}+\cdots\big)\nn\\
&&+\fft{C_4}{r_0^{6}}\big(\frac{2}{r^{10}}+\frac{75 r_0^{4}}{16 r^{14}}-\frac{8 r_0^{6}}{5 r^{16}}+\cdots\big)
+\fft{C_6}{r_0^{10}} \big(\frac{1}{r^6}-\frac{3 r_0^4}{r^{10}}-\frac{225 r_0^8}{32 r^{14}}+\cdots\big)\nn\\
&&+\fft{C_{10}}{r_0^{6}} \big(\frac{1}{r^{10}}+\frac{75 r_0^4}{32 r^{14}}-\frac{4 r_0^6}{5 r^{16}}+\cdots\big)\\
\delta_{q^4}\tilde{h} &=&  \big(-\frac{88}{5625 r^{16}}+\frac{176}{16875 r^{10} r_0^6}+\frac{2251}{78750 r^{14} r_0^2}+\cdots\big)\nn\\
&&+\fft{C_4}{r_0^{12}} \big(\frac{1}{r^4}+\frac{3 r_0^4}{r^8}-\frac{8 r_0^6}{5 r^{10}}+\frac{9 r_0^8}{r^{12}}-\frac{777 r_0^{10}}{80 r^{14}}+\cdots\big)\nn\\
		&&+\fft{C_6}{r_0^{10}} \big(\frac{1}{r^6}+\frac{12 r_0^4}{5 r^{10}}-\frac{2 r_0^6}{r^{12}}+\frac{891 r_0^8}{160 r^{14}}+\cdots\big)
+\fft{C_{10}}{r_0^{6}} \big(\frac{1}{5 r^{10}}+\frac{183 r_0^4}{160 r^{14}}-\frac{4 r_0^6}{5 r^{16}}+\cdots\big)\,, \nn\\
\delta_{q^4}\tilde{\omega} &=&\big(-\frac{83}{10125 \sqrt{6} r^6 r_0^{11}}+\frac{22 \sqrt{\frac{2}{3}}}{1125 r^{10} r_0^7}-\frac{1267 \sqrt{\frac{2}{3}}}{10125 r^{12} r_0^5}+\cdots\big)\nn\\
		&&+\fft{C_4}{r_0^{11}} \big(-\frac{7}{3 \sqrt{6} r^6}+\frac{7 \sqrt{\frac{2}{3}} r_0^6}{3 r^{12}}+\frac{\sqrt{6} r_0^{10}}{5 r^{16}}+\cdots\big)
+\fft{C_6}{r_0^{11}} \big(\frac{7}{2 \sqrt{6} r^6}-\frac{7 r_0^6}{\sqrt{6} r^{12}}-\frac{3 \sqrt{\frac{3}{2}} r_0^{10}}{5 r^{16}}+\cdots\big)\nn\\
		&&+\fft{C_{10}}{r_0^{11}} \big(-\frac{\sqrt{\frac{2}{3}}}{3 r^6}+\frac{2 \sqrt{\frac{2}{3}} r_0^6}{3 r^{12}}+\frac{\sqrt{\frac{3}{2}} r_0^{10}}{5 r^{16}}+\cdots\big)\,.
\end{eqnarray}
At the first sight, the asymptotic structure contain three independent parameters, which would violate the no-hair theorem. However, we shall see from the horizon geometry that for the solution to form a black hole, the three parameters must be completely fixed. Nevertheless, these asymptotic behaviors allow us to read off the conserved quantities such as the mass, angular momentum and the charge. We shall do this presently.

\subsection{Near-horizon geometry}

In our perturbative approach, we fix the horizon radius $r_+=r_0$ unperturbed. In the near-horizon region, we can also define $\delta_{q^4}f = (r-r_0)^\lambda \hat{f}$, where $\hat{f}$ has the usual analytical Taylor expansions at neighborhood near $r\rightarrow r_0$. The leading order behavior of \eqref{f-eom-q4} on the horizon becomes
\begin{equation}
		225(r-r_0)^\lambda r_0^{16} (1+\lambda)(2+\lambda)(\lambda - \Delta_+)(\lambda - \Delta_-) \hat{f}= -128\,,
\end{equation}
where
\begin{equation}
\Delta_\pm = \frac{-3\pm\sqrt{21}}{2} \,. \label{Delta-7D}
\end{equation}
Therefore the general solution near $r_0$ has also four integration constants, taking the form
\begin{equation}
\delta_{q^4}f = \frac{64}{675 r_0^{16}}\hat{f}_0 + \frac{d_{-1}}{r-r_0}\hat{f}_{-1}+ \frac{d_{-2}}{(r-r_0)^{2}}\hat{f}_{-2}+ d_{\Delta_-}(r-r_0)^{\Delta_-}\hat{f}_{\Delta_-}+ d_{\Delta_+}(r-r_0)^{\Delta_+}\hat{f}_{\Delta_+}\,.
\end{equation}
Note that we have $\Delta_+>0$ and $\Delta_-<0$. The regularity at $r_0$ requires that we set the three coefficients $d_{-1}, d_{-2}$ and $d_{\Delta_-}$ all to zero, giving rise to
\be
\delta_{q^4}f = \frac{64}{675 r_0^{16}} \Big(\tilde{f}_{\text{Nor}}+\log\left(\frac{r}{r_0}-1\right) \tilde{f}_{\text{Log}} \Big)+ d_{\Delta_+}(r-r_0)^{\Delta_+}\hat{f}_{\Delta_+}\,,
\ee
where
\begin{eqnarray}
		\hat{f}_{\text{Log}} &=&-\frac{r-r_0}{3 r_0}+\frac{25 \big(r-r_0\big){}^2}{18 r_0^2}-\frac{94 \big(r-r_0\big){}^3}{27 r_0^3}+\cdots\\
		\hat{f}_{\text{Nor}} &=& 1-\frac{\big(r-r_0\big) (-3784+80 \log (2)+\log (3))}{240 r_0}\nn\\
		&&+\frac{\big(r-r_0\big){}^2 (2728+25(80\log (2)+ \log (3)))}{1440 r_0^2}\nn\\
		&&-\frac{\big(r-r_0\big){}^3 (1909+47(80\log (2)+\log (3)))}{1080 r_0^3}+\cdots\,,\nn\\
\hat{f}_{\Delta_+} &=& 1+\frac{117-7 \sqrt{21}}{180r_0}\big(r-r_0\big)+\frac{6627-1282 \sqrt{21} }{18360 r_0^2}\big(r-r_0\big){}^2\nn\\
		&&+\frac{13857-2959 \sqrt{21} }{44064 r_0^3}\big(r-r_0\big){}^3+\cdots\,.
\end{eqnarray}
The logarithmic terms emerge from the Maxwell source \eqref{q4maxsource}. Thus, we see, for given $r_0$, that the horizon geometry involves only one free parameter $d_{\Delta_+}$. For a generic coefficient $d_{\Delta_+}$, it will excite the $C_{-2}$ at the asymptotic region. We thus need to fine-tune this parameter precisely so that $C_{-2}$ vanishes. Thus, for the solution to describe a black hole, all the ``free'' parameters $(C_4,C_6,C_{10})$ and $d_{\Delta_+}$ are completely fixed. We shall determine these coefficients numerically later. Before doing that, we study the possible corrections to the black hole thermodynamics at the NNLO.

\subsection{Corrections to the thermodynamic quantities}

By the standard method, we can read off the mass, charge and angular momentum up to and including the $q^4$ order, based on the asymptotic falloffs of our solution. These are
\begin{eqnarray}
M &=& \frac{15\pi^2}{16} r_0^4-\frac{ \pi ^2 }{24 r_0^4} q^2-\frac{5 \pi ^2 C_4}{16 r_0^{12}} q^4\,,\qquad  Q=\frac{5\pi^2}{8}q\,,\nn\\
J &=& \frac{3}{4}\sqrt{\frac{3}{2}} \pi ^2 r_0^5-\frac{\pi ^2}{4 \sqrt{6} r_0^3} q^2 -\frac{  \Big(166+3375  \big(14 C_1 -21 C_{6}+4 C_{10} \big)\Big)\pi ^2}{54000 \sqrt{6} r_0^{11}} q^4\,.
\label{MQJ}
\end{eqnarray}
Note that $Q$ receives no higher-order correction by construction since it is the perturbation parameter. This leads to the mass/charge relation up to and including the $Q^4$ order, namely
\bea
M &=& \eta_0\Big(J^{\fft45} + \eta_2\, Q^2 J^{-\fft45} + \eta_4\, Q^4 J^{-\fft{12}5}\Big) + {\cal O}(Q^6)\,,\nn\\
\eta_0 &=& \frac{5 \pi ^{2/5}}{4 \sqrt[5]{3}}\,,\qquad
\eta_2 = \frac{2\ 3^{2/5}}{5 \pi ^{4/5}}\,,\qquad
\eta_4 = -\frac{3375 \left(C_4+21 C_6-4 C_{10}\right)+284}{625 \sqrt[5]{3} \pi ^{8/5}}\,.\label{etas}
\eea
We shall determine the coefficient $\eta_4$ by a numerical method presently. On the horizon, we can read off the entropy, electric potential and angular velocity
\bea
S &=& \frac{1}{4} \pi ^3 \sqrt{3} r_0^5 -\frac{\pi ^3 q^2}{12 \sqrt{3} r_0^3}
-\frac{5 \pi ^3 q^4}{216 \sqrt{3} r_0^{11}}\,,\qquad \Phi_H = \frac{1}{3 r_0^4} q +\frac{32}{675 r_0^{12}} q^3\,,\\
\Omega_H &=& \frac{\sqrt{\frac{2}{3}}}{r_0}-\frac{ \sqrt{\frac{2}{3}}}{45 r_0^9} q^2+ \Big(\delta_{q^4}\omega\big(r_0\big)+\frac{8\sqrt{\frac{2}{3}}}{81 r_0^{17}}\Big)q^4\,.
\eea
In this derivation, we have made use of the fact that $\delta_{q^4}W\big(r_0\big) = -\frac{4}{27r_0^{16}}$ from the first equation in \eqref{eom on omega}. However, the explicit form of $\delta_{q^4} \omega\big(r_0\big)$ remains to be determined. The entropy can be expressed as a function of angular momentum and the charge
\be
S = \frac{\sqrt{2} \pi }{3} (J  + \xi_4 Q^4 J^{-\fft{11}5}\Big) + {\cal O}(Q^6)\,,\qquad
\xi_4 =  \frac{3375 \left(14 C_4-21 C_6+4 C_{10}\right)-3584}{500 \sqrt[5]{3} \pi ^{8/5}}\,.
\label{xis}
\ee
The first law of black hole thermodynamics at zero temperature ($TdS=0$) is given by \eqref{firstlaw}, which provides nontrivial constraints on the parameters of the perturbative solutions. Specifically, we must have
\begin{eqnarray}
0&=&\frac{\pi ^2 q^4}{162000 r_0^{13}} \Big(60926 - 3375 (26 C_4 + 231 C_6 -44 C_{10} ) + 303750\sqrt6 r_0^{17} \delta_{q^4}\omega(r_0)\Big) dr_0\nn\\
&&+\frac{\pi ^2 q^3}{40500 r_0^{12}}\Big(464+3375(C_4 + 21 C_6 - 4 C_{10})\Big)dq\,.\label{firstlawtest}
\end{eqnarray}
In the next subsection, we shall perform explicit numerical calculation and determines the
parameters $(C_4,C_6,C_{10})$ and $r_0^{17} \delta_{q^4}\omega(r_0)$ completely. The correct numerical solution describing a black hole should give a vanishing result of the above equation.

\subsection{Numerical analysis and results}
	
\begin{figure}[ht]
	\centering
	\includegraphics[width=0.45\linewidth]{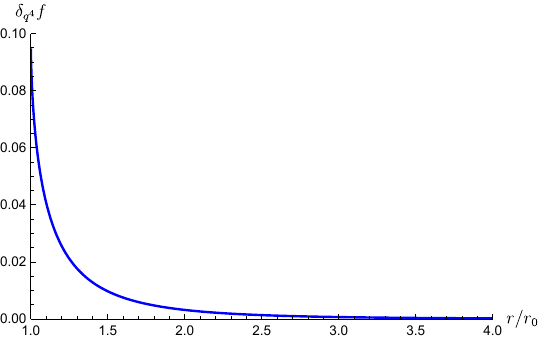} \includegraphics[width=0.45\linewidth]{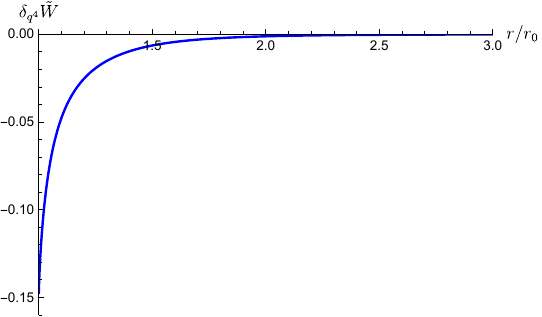}
\includegraphics[width=0.45\linewidth]{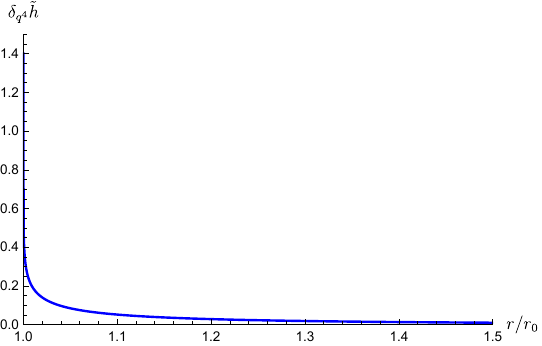} \includegraphics[width=0.45\linewidth]{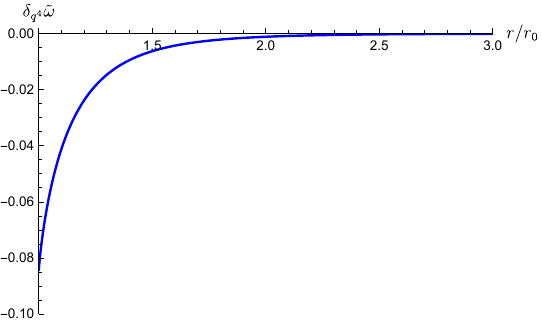}
	\caption{\small Here are the four $\delta_{q^4}$ metric functions, with the integration constants chosen so that the functions decay at large $r$.}
	\label{fig1}
\end{figure}

We use the standard shooting method to numerically solve all the $\delta_{q^4}$ perturbations. First we work in $x \equiv \frac{r-r_0}{r+L}$ representation, where $r\in (r_0, \infty) \Longleftrightarrow  x\in (0,1)$, so that we can strictly define asymptotic infinity at $x=1$ in the numerical calculations. The value of $L$ does not affect the boundary points or the numerical results, but a suitable choice of $L$ can make computations near the boundary more efficient. For example, a large $L$ can push the corresponding $r$ of given $x$ (e.g.~$x=0.99$) further out to infinity. We use the power-series expansion of the near-horizon geometry as the boundary and numerically integrate out to infinity. In Fig.~\ref{fig1}, we plot all the four $\delta_{q^4}$ metric functions, based on our numerical calculations.

  By shooting method, we mean that we choose a suitable $d_{\Delta_+}$ coefficient so that the
function $\delta_{q^4} f$ vanishes at $x\rightarrow 1$. Having obtained the numerical data, we can perform the curving fitting of the data at the asymptotic region with the large-$r$ power-series structures, and read off the coefficients $(C_4,C_6,C_{10}$). We find
\begin{table}[ht]
		\centering
		\begin{tabular}{|c| c| c| c| }
			\hline
			$d_{\Delta_+}$ & $C_{4}$ & $C_{6}$ & $C_{10}$  \\
			\hline
			$-1.2327$ & $0.061630$ & $-0.075285$ & $-0.34547$ \\
			\hline
		\end{tabular}
		\caption{\small Here are the parameters of the perturbation $\delta_{q^4} f$ from the numerical calculation.}
	\end{table}

Our numerical analysis indicates that the vanishing of $\delta_{q^4} \omega$ implies that $r_0^{16}\delta_{q^4}\omega(r_0) = -0.084553$. We can now verify the first law \eqref{firstlaw} by checking whether the equation \eqref{firstlawtest} vanishes.  It is easy to verify that our numerical results fit the two constraints up to $0.1\%$ precision. This precision strongly indicates that our numerical solution is indeed a black hole and that the irrational exponent $\Delta_+$ can indeed exist in the near-horizon power-series expansion.

Substituting these numerical data to \eqref{etas} and \eqref{xis}, we have
\be
\eta_4 \sim 0.0370233\,,\qquad \xi_4 = -8.03556\times 10^{-7}\sim 0\,.
\ee
The fact that $\xi_4\sim 0$ suggests we should simply have $\xi_4=0$. Together the constraints from the first law, we have three equations on the four parameters, $(C_4,C_6,C_{10})$ and $\delta_{q^4}\omega(r_0)$. Together with our numerical data, we rationalize these coefficients and give
\be
C_4=\frac{208}{3375},\qquad C_6=-\frac{254}{3375},\qquad C_{10}=-\frac{259}{750}\,,\qquad
\sqrt{6} r_0^{16} \delta_{q^4} \omega(r_0) = - \fft{233}{1125}\,.
\ee
Consequently, from \eqref{etas}, we have
\be
\eta_4=\frac{12\ 3^{4/5}}{125 \pi ^{8/5}}\,.
\ee
It is easy to see that these precise numbers fit the numerical data in high accuracy.

\section{Conclusions}

In this paper, we studied the small charge perturbation to the (leading-order) extremal MP solutions with all equal angular momenta in EM gravity in odd $D=2n+1$ dimensions. Analytical solutions could be obtained at the NLO, which allowed us to obtain the small-charge corrections to the black hole thermodynamic quantities of the MP solutions. In particular, we found from \eqref{NLOmasscharge} that the NLO contributed positively to the mass.

The NNLO solution cannot be fully solved analytically and we used $D=7$ dimensions as an illustrative example. We found that irrational exponent $\Delta_+$ in the near-horizon geometry, discovered first in the higher-derivative perturbation \cite{Mao:2023qxq}, also emerged, as predicted in \cite{Mao:2023qxq}. There are two important aspects that are worth emphasizing. First, the Maxwell field does not have the irrational exponent at the NNLO and it is not difficult to envision that such irrational exponent will emerge in the Maxwell field at the higher order. In other words, we cannot blame the matter for the direct emergence of the irrational exponent.  In fact, in this particular example of order-by-order perturbative approach, the matter source, i.e.~the Maxwell field will develop irrational exponents because of the metric having such an exponent first.

The second point that is very important to us is that the existence of the horizon geometry does not necessarily lead to a black hole. We used the numerical analysis, confirmed by the perturbative first law, to show that the horizon geometry with irrational exponents indeed formed a black hole by integrating the horizon geometry to asymptotic Minkowski spacetime. In doing so, a fine-tuning was necessary so that the horizon free parameter $d_{\Delta_+}$ was uniquely fixed. Consequently, the black hole satisfied the no-hair theorem, involving only the mass, angular momentum and charge parameters. From the numerical results, we conjectured the exact perturbative expressions of mass and entropy as functions of angular momentum and charge, up to and including the $Q^4$ order. Curiously, the entropy receives no corrections at all. The existence of these precise numbers indicates an analytic approach, and it is of great interest to confirm such results with some alternative methods.

\section*{Acknowledgement}

This work is supported in part by the National Natural Science Foundation of China (NSFC) grants No.~12375052 and No.~11935009, and also by the Tianjin University Self-Innovation Fund Extreme Basic Research Project Grant No.~2025XJ21-0007.

\appendix
\section{Nonlinear differential equations of motion}
\label{app:gen-eom}

In Section 2, we presented the ansatz \eqref{metric-unpertur-D} for charged rotating black holes in $D=2n+1$ dimensions with all equal angular momenta. Substituting the ansatz into the equations of motion leads to six coupled nonlinear ordinary differential equations:
\begin{eqnarray}
		0 &=& -h(r) \Big((6-2 D) r^2 W(r)^3+(6-2 D) W(r)^2 \Big((1-D) r^2+(-2+D) r^2 f(r)\nn\\
		&&+\Phi (r)^2\Big)+r^4 f(r) W'(r)^2+r^2 f(r) W(r) \Big(2 r W'(r)+\Phi '(r)^2\Big)\Big)\nn\\
		&&+r^3 f(r) W(r) \Big((-4+2 D) W(r) h'(r)+r h'(r) W'(r)+r W(r)^2 \Big(-\omega (r) \Phi '(r)\nn\\
		&&+\Psi '(r)\Big)^2+r^3 W(r)^3 \omega '(r)^2\Big),\cr
		0 &=& -2 (-3+D) (-1+D) r^2 h(r) W(r)^2+2 (-3+D) (-2+D) r^2 f(r) h(r) W(r)^2\nn\\
		&&+2 (-3+D) r^2 h(r) W(r)^3+2 (-3+D) h(r) W(r)^2 \Phi (r)^2)\nn\\
		&&+2 (-2+D) r^3 h(r) W(r)^2 f'(r+2 (-1+D) r^3 f(r) h(r) W(r) W'(r)\nn\\
		&&+r^4 h(r) W(r) f'(r) W'(r)+f(r) \Big(-r^4 h(r) W'(r)^2+r^4 W(r)^3 \Big(-\omega (r) \Phi '(r)\nn\\
		&&+\Psi '(r)\Big)^2+r^6 W(r)^4 \omega '(r)^2+r^2 h(r) W(r) \Big(\Phi '(r)^2+2 r^2 W''(r)\Big)\Big),\cr
		0 &=& h(r) W(r) \Big(-r^2 W(r) f'(r) h'(r)+2 h(r) \Big(-2 (-3+D) W(r)^2+r W(r) f'(r)\nn\\
		&&+r^2 f'(r) W'(r)\Big)\Big)+f(r) \Big(r^2 W(r)^2 h'(r)^2+2 r h(r) W(r) \Big(r h'(r) W'(r)\nn\\
		&&+r W(r)^2 \Big(-\omega (r) \Phi '(r)+\Psi '(r)\Big)^2+2 r^3 W(r)^3 \omega '(r)^2\nn\\
		&&-W(r) \Big((-3+D) h'(r)+r h''(r)\Big)\Big)+2 h(r)^2 \Big(2 (-3+D) W(r)^2-2 r^2 W'(r)^2\nn\\
		&&+W(r) \Big(2 (-2+D) r W'(r)+\Phi '(r)^2+2 r^2 W''(r)\Big)\Big)\Big),\cr
		0 &=& r^2 h(r) W(r) f'(r) \omega '(r)+f(r) \Big(-r^2 W(r) h'(r) \omega '(r)+2 h(r) \Big(-\omega (r) \Phi '(r)^2\nn\\
		&&+\Phi '(r) \Psi '(r)+r \Big(D W(r) \omega '(r)+2 r W'(r) \omega '(r)+r W(r) \omega ''(r)\Big)\Big)\Big),\cr
		0 &=& r h(r) W(r) f'(r) \Big(\omega (r) \Phi '(r)-\Psi '(r)\Big)+f(r) \Big(r W(r) h'(r) \Big(-\omega (r) \Phi '(r)\nn\\
		&&+\Psi '(r)\Big)+2 h(r) \Big(r W'(r) \Big(\omega (r) \Phi '(r)-\Psi '(r)\Big)+W(r) \Big(-\Big((-2+D) \Psi '(r)\Big)\nn\\
		&&+\omega (r) \Big((-2+D) \Phi '(r)+r \Phi ''(r)\Big)+r \Big(\Phi '(r) \omega '(r)-\Psi ''(r)\Big)\Big)\Big)\Big),\cr
		0&=&r^4 f(r) W(r)^3 \omega (r) h'(r) \Big(\omega (r) \Phi '(r)-\Psi '(r)\Big)\nn\\
		&&+h(r)^2 \Big(-4 (-3+D) W(r)^2 \Phi (r)-2 r^2 f(r) W'(r) \Phi '(r)+r W(r) \Big(r f'(r) \Phi '(r)\nn\\
		&&+2 f(r) \Big((-4+D) \Phi '(r)+r \Phi ''(r)\Big)\Big)\Big)+h(r) \Big(r^4 W(r)^3 \omega (r) f'(r) \Big(-\omega (r) \Phi '(r)\nn\\
		&&+\Psi '(r)\Big)+r^2 f(r) W(r) \Big(h'(r) \Phi '(r)+2 r^2 W(r) \omega (r) W'(r) \Big(-\omega (r) \Phi '(r)\nn\\
		&&+\Psi '(r)\Big)-2 r W(r)^2 \Big(-r \Psi '(r) \omega '(r)+\omega (r)^2 \Big((-2+D) \Phi '(r)+r \Phi ''(r)\Big)\nn\\
		&&-\omega (r) \Big((-2+D) \Psi '(r)+r \Big(-2 \Phi '(r) \omega '(r)+\Psi ''(r)\Big)\Big)\Big)\Big)\Big).\label{gen-eom}
	\end{eqnarray}

\end{document}